\documentclass[11pt]{article}

\usepackage[margin=1in]{geometry}
\usepackage{lmodern}
\usepackage{sidecap}
\usepackage{xspace}
\usepackage{doi}
\usepackage{hyperref} 
\usepackage{tikz}
\usepackage{pgfgantt}
\usepackage{graphicx,pbox}
\usepackage{xcolor,hyperref,colortbl}
\usepackage{wrapfig}
\usepackage{enumitem}
\usepackage{gensymb}
\usepackage{comment}
\usepackage{lineno}
\newcommand\pubnumber{}
\newcommand\pubdate{\today}
\newcommand\pubblock{\rightline{\begin{tabular}{l} \pubnumber\\
          \pubdate \end{tabular}}}

\def\Title#1{\begin{center} {\Large #1 } \end{center}}
\def\Author#1{\begin{center}{ \sc #1} \end{center}}

\newcommand\snowmass{\begin{center}\rule[-0.2in]{\hsize}{0.01in}\\\rule{\hsize}{0.01in}\\
\vskip 0.1in Submitted to the Oxford Encylopedia of Physics\\  
\rule{\hsize}{0.01in}\\\rule[+0.2in]{\hsize}{0.01in} \end{center}}
\begin{document}
\begin{titlepage}
\snowmass
\pubblock

\Title{\huge {\bf Ultimate Colliders} $^\dagger$
}

\Author{Vladimir D. Shiltsev $^1$ \\
Fermi National Accelerator Laboratory, Batavia, IL 60510, USA } 
{\bf Summary:}

Understanding the Universe critically depends on the fundamental knowledge of particles and fields, which represents a central endeavor of modern high-energy physics. Energy frontier particle colliders -- arguably, among the largest, most complex and advanced scientific instruments of modern times -- for many decades have been at the forefront of scientific discoveries in high-energy physics. Due to technology advances and beam physics breakthroughs, the colliding beam facilities have progressed immensely and now operate at energies and luminosities many orders of magnitude greater than the pioneering instruments of the early 1960s. 

While the Large Hadron Collider and the Super-KEKB factory represent the frontier hadron and lepton colliders of today, respectively, future colliders are an essential component of a strategic vision for particle physics.  Conceptual studies and technical developments for several exciting near- and medium-term future collider options are underway internationally. Analysis of numerous proposals and studies for far-future colliders indicate the limits of the collider beam technology due to machine size, cost, and power consumption, and call for a paradigm shift of particle physics research at ultra-high energy but low luminosity colliders approaching or exceeding 1 PeV center-of-mass energy scale.   
 \\

\\

\noindent
{\bf Subjects:} High Energy Physics, Particle Accelerators
\\
\\
\noindent
$^1${email: shiltsev@fnal.gov } \\
\\
$^{\dagger}$ This work has been supported by the Fermi Research Alliance, LLC under Contract No. DE-AC02-07CH11359 with the U.S. Department of Energy, Office of Science, Office of High Energy Physics. \\ 
\end{titlepage}

\newpage
\tableofcontents
\newpage

\section{Introduction} 
\label{intro}

Particle accelerators are unique scientific instruments which offer access to unprecedented energy per constituent, using well-focused, high-density beams of electrons ($e^-$), positrons ($e^+$), protons ($p$), antiprotons ($\bar {p}$), ions, muons ($\mu^+$, $\mu^-$), mesons, photons, and gamma quanta ($\gamma$), among others \cite{ShiltsevPhysicsToday}. Three Nobel prizes were awarded for seminal advancements in accelerator science and technology: to Ernest O.~Lawrence in 1939 for invention of the first modern accelerator, the cyclotron \cite{lawrence1932}, to John Cockcroft and Ernest Walton in 1951 for their invention of the eponymous linear accelerator \cite{cockcroft}, and to Simon van der Meer in 1984 for conceiving and developing the novel method of stochastic cooling \cite{vandermeer1985}. Of course, highly notable are applications of accelerators - for example, they were of critical importance for about a quarter of the most acclaimed physics discoveries since 1939, resulting on average in a 
Nobel Prize for Physics every three years \cite{HC}. Electron microscopes, accelerator-based synchrotron radiation and spallation neutron sources were instrumental for numerous Nobel Prize-winning research achievements in chemistry, physiology and medicine, such as those recognized in 1997, 2003, 2006, 2009, 2012, 2017, 2019, and 2021. 
 
\begin{figure}[htbp]
\centering
\includegraphics[width=0.99\linewidth]{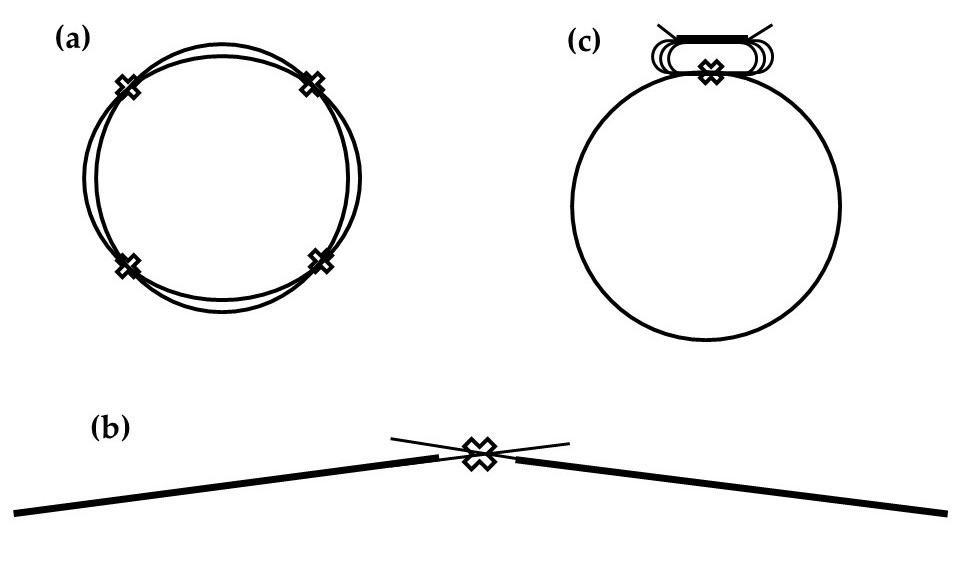}
\caption{Schematics of some particle collider types: a) circular, b) linear, c) ring-ERL(energy recovery linac). Beam collision points are marked by crosses.} \label{fig:types}
\end{figure}

At present, about 140 accelerators of all types worldwide are devoted to fundamental research \cite{faus2017applications}. Among them, the most complex and technologically advanced are higher-energy accelerators and, especially, colliders for nuclear and particle physics. While they are of different sizes and shapes, based on different technologies and employing different types of particles, they have common functional elements and basic stages -- charged particles are produced in dedicated sources, often go through a preparatory stage to arrange the particles in suitable beams of bunches, and then get accelerated to very high kinetic energies. (Here it is generally assumed that all the particles are ultra-relativistic and their kinetic energy and full energy are the same $E=\gamma m c^2$, where $m$ is the particle's mass, $c$ is the speed of light, and relativistic Lorentz factor $\gamma \gg 1$.) In order to be most effective in getting insights into the interesting physics of nuclei and/or elementary particles, the beams usually get compressed in a sequence of dedicated elements, like focusing magnets, before being sent to strike other particles, causing reactions that transform the particles into new particles. Sophisticated detectors are needed to identify and analyse products of the reactions of interest. 

What makes colliders distinct is the use of two similar but counter-propagating beams directed onto each other in one or several interaction points (IPs) -- see Fig.\ref{fig:types}. While such an arrangement makes the machines significantly more complex \cite{shiltsev2021modern}, it is fully justified  by the enormous kinematic advantage in so-called center-of-mass energy, resulting in much larger available energy and, therefore, opportunity to generate new particles of much higher masses. 
Indeed, for the head-on collision of two ultra-relativistic particles with equal energy $E$, the center of mass energy (c.m.e.) is:   
\begin{equation}
E_{cm}  \approx  2E .
\label{E1}
\end{equation}
\noindent (The equation for unequal particle energies $E_1 \neq E_2$ is $E_{cm} \approx 2\sqrt{E_1 E_2}$). High-energy particles can also be sent onto a stationary target, resulting in $E_{cm} \approx \sqrt{2 E mc^2}$, where $m$ is the mass of the target-material particles. Take, for example, the highest energy cosmic rays observed on Earth, reaching $E \sim 10^{21}$ eV, or a million PeV (1 PeV=1000 TeV=1000,000 GeV=$10^{15}$ eV). Their collisions with stationary protons ($mc^2 \approx 1$ GeV) result in the c.m.e. of 1.4 PeV. In comparison, the same c.m.e. would be possible in a particle collider with only $E$=0.7 PeV=700 TeV energy per beam, i.e., with a million(!) times smaller particle energies. The highest beam and center-of-mass energies achieved to date are, of course, much lower - $E$=0.007 PeV and $E_{cm}$=0.014 PeV in the Large Hadron Collider (LHC), see Table \ref{T1}. In what follows, the ultimate limits of particle colliders are discussed. 

\begin{table}
\begin{tabular}{|l|c|c|c|c|c|l|}
\hline
Colliders & Species & $E_{cm}$, GeV & $C$, m & ${\cal L}$, $10^{32}$ & Years & Host lab, country \\
\hline
AdA & $e^+e^-$ & 0.5 & 4.1 & $10^{-7}$  & 1964 & Frascati/Orsay\\
VEP-1 & $e^-e^-$ & 0.32 & 2.7 & $5\times10^{-5}$  & 1964-68 & Novosibirsk, USSR \\
CBX & $e^-e^-$ & 1.0 & 11.8 &$ 2\times10^{-4}$  & 1965-68 & Stanford, USA \\
VEPP-2 & $e^+e^-$ & 1.34& 11.5 &$ 4 \times10^{-4}$  & 1966-70 & Novosibirsk, USSR \\
ACO & $e^+e^-$ & 1.08 & 22 & 0.001  & 1967-72 & Orsay, France \\
ADONE & $e^+e^-$ & 3.0 & 105 & 0.006  & 1969-93 & Frascati, Italy \\
CEA & $e^+e^-$ & 6.0 & 226 & $0.8\times10^{-4}$  & 1971-73 & Cambridge, USA\\
ISR & $pp$ & 62.8 & 943 & 1.4  & 1971-80 & CERN \\
SPEAR & $e^+e^-$ & 8.4 & 234 & 0.12  & 1972-90 & SLAC, USA\\
DORIS & $e^+e^-$ & 11.2 & 289 & 0.33  & 1973-93 & DESY, Germany \\
VEPP-2M & $e^+e^-$ & 1.4 & 18 & 0.05  & 1974-2000 & Novosibirsk, USSR \\
VEPP-3 & $e^+e^-$ & 3.1 & 74 & $2\times10^{-5}$  & 1974-75 & Novosibirsk, USSR \\
DCI & $e^+e^-$ & 3.6 & 94.6 & 0.02  & 1977-84 & Orsay, France \\
PETRA & $e^+e^-$ & 46.8 & 2304 & 0.24  & 1978-86 & DESY, Germany \\
CESR & $e^+e^-$ & 12 & 768 & 13  & 1979-2008 & Cornell, USA\\
PEP & $e^+e^-$ & 30 & 2200 & 0.6  & 1980-90 & SLAC, USA \\
S$p\bar{p}$S & $p \bar{p}$ & 910 & 6911 & 0.06  & 1981-90 & CERN \\
TRISTAN & $e^+e^-$ & 64 & 3018 & 0.4  & 1987-95 & KEK, Japan \\
Tevatron & $p \bar{p}$ & 1960 & 6283 & 4.3  & 1987-2011 & Fermilab, USA\\
SLC & $e^+e^-$ & 100 & 2920 & 0.025 & 1989-98 & SLAC, USA\\
LEP & $e^+e^-$ & 209.2 & 26659 & 1  & 1989-2000 & CERN \\
HERA & $ep$ & 30+920& 6336 & 0.75  & 1992-2007 & DESY, Germany \\
PEP-II & $e^+e^-$ & 3.1+9 & 2200 & 120  & 1999-2008 & SLAC, USA\\
KEKB & $e^+e^-$ & 3.5+8.0 & 3016 & 210  & 1999-2010 & KEK, Japan\\
\hline
VEPP-4M & $e^+e^-$ & 12 & 366 & 0.22  & 1979- & Novosibirsk, Russia\\
BEPC-I/II & $e^+e^-$ & 4.6 & 238 & 10  & 1989- & IHEP, China\\
DA$\Phi$NE & $e^+e^-$ & 1.02 & 98 & 4.5  & 1997- & Frascati, Italy \\
RHIC & $p,i$ & 510 & 3834 & 2.5  & 2000- & BNL, USA \\
LHC & $p,i$ & 13600 & 26659 & 210  & 2009- & CERN \\
VEPP2000 & $e^+e^-$ & 2.0 & 24 & 0.4  & 2010- & Novosibirsk, Russia\\
S-KEKB & $e^+e^-$ & 7+4 & 3016 & 6000$^*$ & 2018- & KEK, Japan \\
\hline
NICA & $p,i$ & 13 & 503 & 1$^*$  & 2024(tbd) & JINR, Russia \\
EIC & $ep$ & 10+275& 3834 & 105$^*$  & 2032(tbd) & BNL, USA\\
\hline
\hline
Proposals & Species & $E_{cm}$, TeV & $C$, km & ${\cal L}^{*}$, $10^{35}$ & Years & Host lab, country\\
\hline
FCCee & $e^+e^-$ & 0.24 & 91 & 0.5  & n/a & CERN \\
CEPC & $e^+e^-$ & 0.24 & 100 & 0.5  & n/a & China \\
ILC-0.25 & $e^+e^-$ & 0.25 & 20.5 & 0.14 & n/a & Japan \\
CLIC-0.38 & $e^+e^-$ & 0.38 & 11 & 0.15  & n/a & CERN \\
ILC-1 & $e^+e^-$ & 1 & 38 & 0.5 & n/a & Japan \\
LHeC & $ep$ & 0.06+7& 9+26.7 & 0.08  & n/a & CERN \\
CLIC-3 & $e^+e^-$ & 3 & 50 & 0.6  & n/a & CERN \\
MC-3 & $\mu^+\mu^-$ & 3 & 4.5 & 0.18 & n/a & n/a \\
MC-14 & $\mu^+\mu^-$ & 14 & 14 & 4  & n/a & n/a \\
WFA-15 & $e^+e^-$ & 15 & ~12 & 5 & n/a & n/a \\
WFA-30 & $e^+e^-$ & 30 & ~20 & 32  & n/a & n/a \\
FCChh & $pp$ & 100 & 91 & 3 & n/a & CERN \\
SPPC & $pp$ & 125 & 100 & 1.3 & n/a & IHEP, China \\
\hline 
\end{tabular}
\caption{Past, present and several proposed future particle colliders: their particle species, center of mass energy $E_{cm}$, circumference or length $C$, maximum peak luminosity ${\cal L}$ per interaction point, years of luminosity operation, and host labs. ($i$ is for ions; luminosity is in units of cm$^{-2}$s$^{-1}$, $^*$ design; see also text.)} 
\label{T1}
\end{table}

\section{Colliders: Energy, Luminosity, History}

As noted above, colliders essentially shaped modern particle physics, and 31 of them have so far reached the operational stage  (some in several successive configurations), with seven operational now (2023) -- see Table \ref{T1}. Two colliders are under construction and almost three dozen proposals for future colliders are under discussion, some of which are also listed in Table \ref{T1}. The idea of using colliding beams to gain the above mentioned kinematic advantage was first given serious consideration by the Norwegian engineer and inventor Rolf Wider\"{o}e, who in 1943 had filed a patent for the collider concept (and received the patent in 1953) \cite{Wideroe, waloschek2013}, and then further developed by Donald Kerst \cite{kerst} and Gerry O\textsc{\char13}Neill \cite{o'neill}. 
In the early 1960s, almost concurrently, three early colliders went  into operation in the Soviet Union ($e^-e^-$ collider VEP-1), France (to where the $e^+e^-$ AdA had been moved from Italy), and the USA ($e^-e^-$ CBX). 

The first colliders, as well as all but one follow up machine, were built in a storage ring (circular) configuration -- see Fig.~\ref{fig:types}a -- where particles of each beam circulate in the same or two different rings and repeatedly collide. In linear colliders, first proposed in Ref.~\cite{tigner1965} and realized in the 1990s in the SLAC Linear Collider (SLC), the two colliding beams are accelerated in linear accelerators (linacs) and transported to a collision point, either in a simple two-linac configuration as depicted in Fig.~\ref{fig:types}c or with use of the same linac and two arcs, as in the SLC. Other configurations are possible and were considered: e.g., collision of beams circulating in a ring and a few-pass energy recovery linac (ERL) (Fig.~\ref{fig:types}b) or linac-ring schemes.   

\begin{figure}[htbp]
\centering
\includegraphics[width=0.99\linewidth]{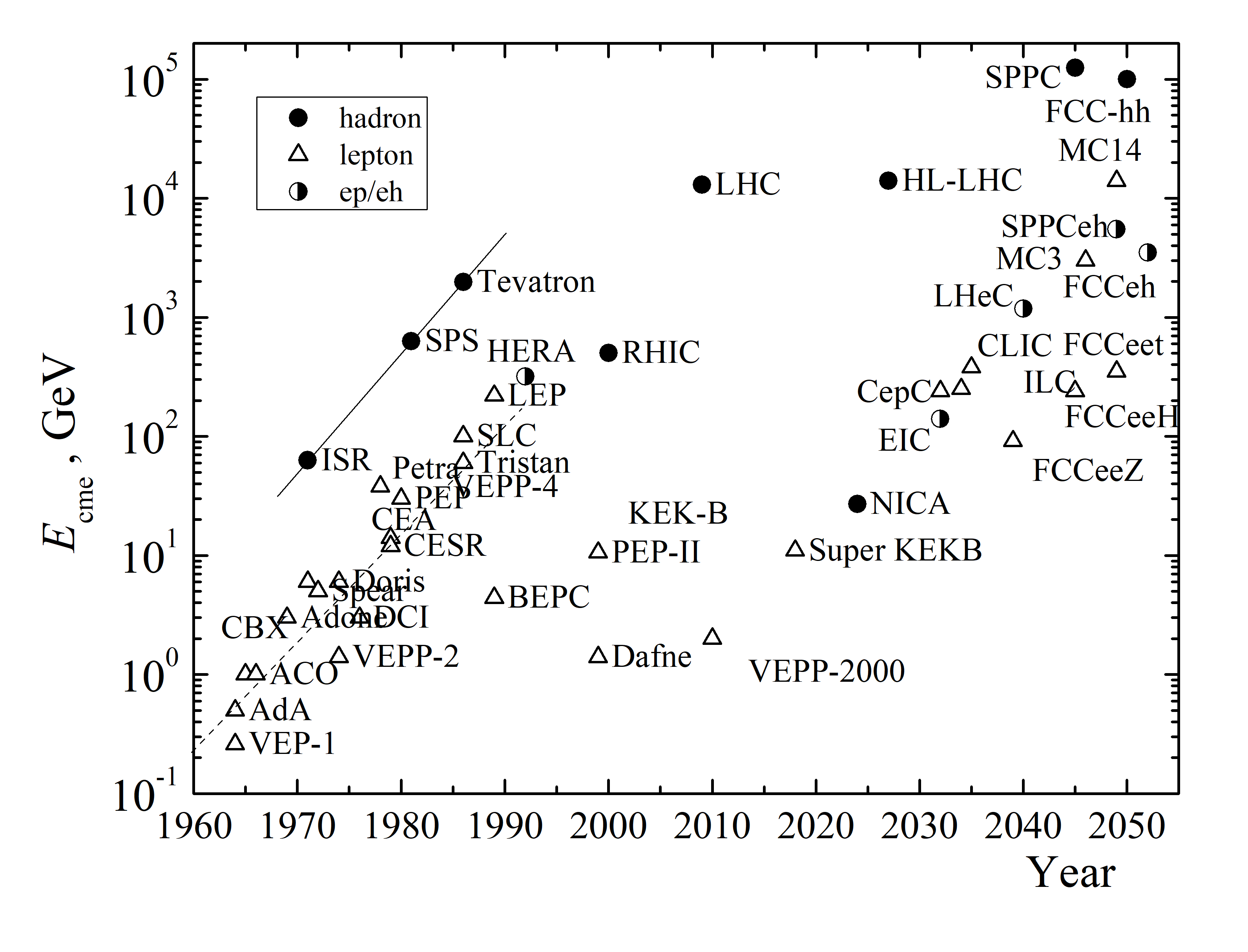}
\caption{Center of mass energy reach of particle colliders vs their actual or proposed start of operation. Solid and dashed lines indicate a ten-fold increase per decade for hadron (circles), lepton (triangles) and lepton-hadron (half filled circles) colliders (adapted from \cite{shiltsev2021modern}).}
\label{fig:colliders_E}
\end{figure}

The ever-growing demands of particle physics research drove an increase in the beam energy and c.m.e. of colliders by five orders of magnitude, as is demonstrated in Fig.~\ref{fig:colliders_E}. 
Charged particles gain energy from an electric field. The accelerating-field gradients in fast time-varying structures, such as radio-frequency (RF) cavities, are  usually orders of magnitude higher than in direct-current (DC) systems, and, therefore, commonly used in modern colliders (with the RF frequencies ranging from 10s of MHz to 10s of GHz). 
At present, the highest beam accelerating gradients ever achieved in operational machines or beam-test facilities are about 31.5 MV/m in 1.3 GHz superconducting RF (SRF) cavities and some $G\approx 100$~MV/m in 12 GHz normal-conducting (NC) ones. The much higher gradients $O(10 GV/m)$ are reached in plasma wake-field acceleration (WFA) experiments (see below). 
In a linear-collider arrangement, illustrated in Fig.~\ref{fig:types}c, the beam energy $E$ is the product of the average accelerating gradient $G$ and the length of the linac $L$:
\begin{equation}
    E = e G \cdot L
    \; ,
\label{eq:energy_l}
\end{equation}
where $e$ denotes the elementary (electron) charge, assuming the acceleration of singly charged particles like electrons or protons.  For example, reaching just 0.001 PeV=1 TeV energy requires either  $\sim$30 km of SRF linac or 10 km of NC RF accelerator, if the RF cavities occupied all available space -- which they usually do not.  

Cost considerations (see below) imply that RF acceleration hardware, such as normally metallic resonant cavities, RF power sources and distribution systems, should be minimized, e.g., through repeated use of the same RF system, which would boost the energy incrementally, $\Delta E=eV_{RF}$ per turn every time a particle passes through the total cavity voltage $V_{RF}$. Such an arrangement can be realized both in the form of circular colliders (Fig.~\ref{fig:types}a), which have proven extremely  successful, and also through schemes based on ERLs  (Fig.~\ref{fig:types}b). 
Circular colliders are most common; here, dipole magnets with an average magnetic field $B$ and bending radius, $\rho$,  are used to confine charged particles inside the accelerator beam pipe passing through the apertures of the dipoles such that : 
\begin{equation}
    E=ec B \cdot \rho \quad \textup{or} \quad E\; \textup{[TeV]} = 0.3 (B\rho)\; \textup{[T $\cdot$ km]} \; . 
\label{eq:energy_c}
\end{equation} 
As the particles are accelerated in a {\it synchrotron}, the strength of the magnetic field is increased to keep the radius of the orbit approximately constant. 

The maximum field of NC magnets is about 2 Tesla (T) due to the saturation of ferromagnetic materials, and while this  is sufficient for lower-energy colliders, such as most  $e^+e^-$ storage rings, it is not adequate for very high-energy hadron or muon beams because it would require excessively long accelerator tunnels and prohibitively high magnet power consumption. The development of superconducting (SC) magnets carrying high electric current in Nb-Ti wires cooled by liquid helium below 5~K opened the way towards higher fields and to hadron colliders at record energies \cite{tollestrup2008}. For example, the 14 TeV c.m.e.~LHC at CERN uses double-bore magnets with a maximum field of 8.3 T at a temperature of 1.9 K 
in a tunnel of $C=26.7$ km circumference (dipole-magnet bending radius $\rho=2,800$~m).

\begin{figure}[htbp]
\centering
\includegraphics[width=0.99\linewidth]{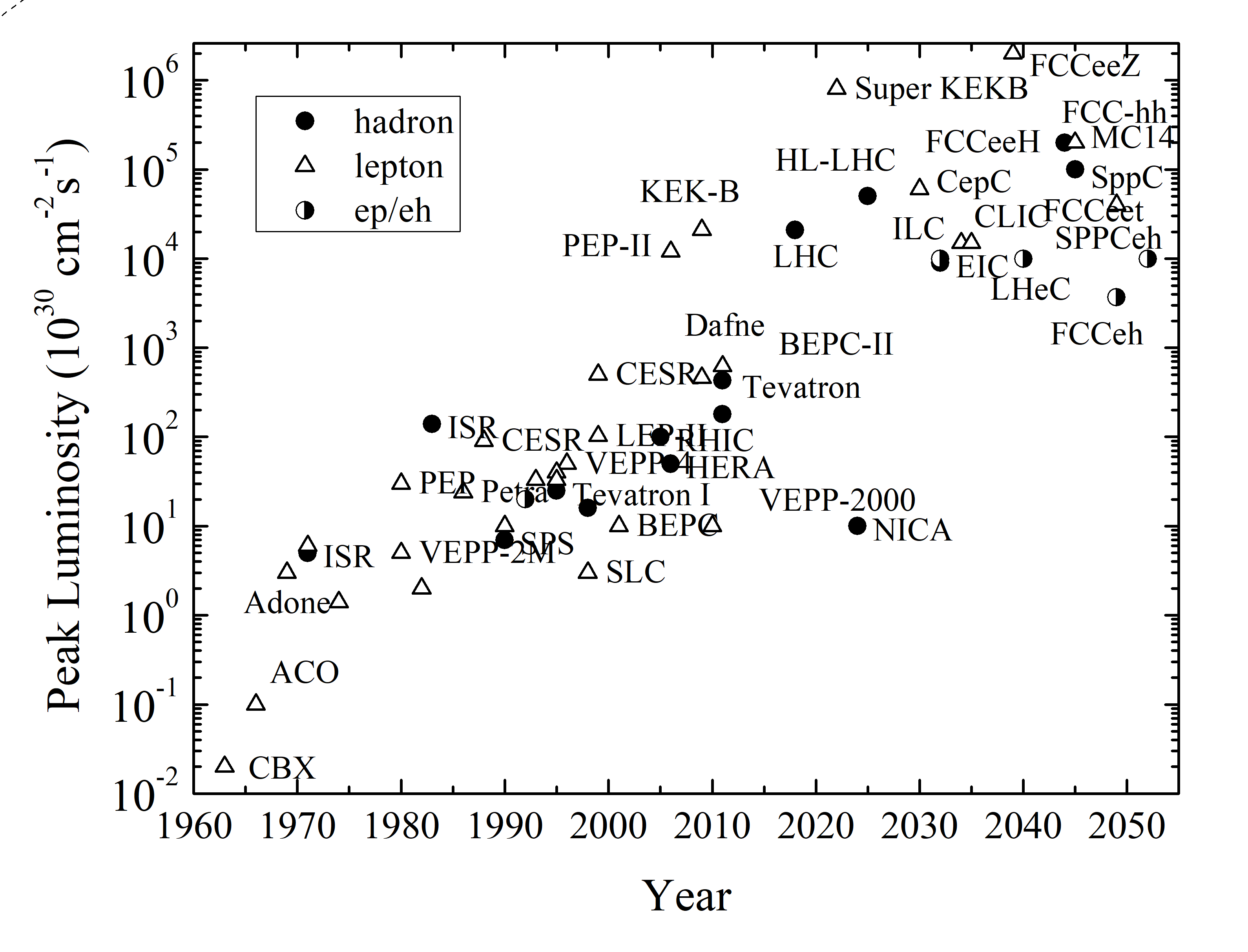}
\caption{Luminosities of particle colliders: triangles are lepton colliders, full circles are hadron colliders, and half-filled circles for electron-hadron colliders. Values are per collision point (adapted from \cite{shiltsev2021modern}). }
\label{fig:colliders_L}
\end{figure}

The exploration of rare particle-physics phenomena at the energy frontier requires not only an appropriately high energy, but also a sufficiently large number of detectable reactions. This number, $N_{reaction}$ is given by  the product of the cross section of the reaction under study, $\sigma_{\textup{reaction}}$, and the time integral over the instantaneous {\it collider luminosity}, $\cal L$:
\begin{equation}
N_{\textup{reaction}} =  \sigma_{\textup{reaction}} \cdot \int {\cal L} (t) dt.
\label{eq:intlumi}
\end{equation}
The luminosity dimension is [length]$^{-2}$[time]$^{-1}$. The integral on the right is referred to as {\it integrated luminosity} ${\cal L}_{\rm int}$, and, reflecting the smallness of typical particle-interaction cross-sections, is often reported in  units of inverse pico-, femto- or attobarns. By definition, 1 barn is equal to 10$^{-24}$~cm$^2$, and, correspondingly, 1 ab$^{-1}$=10$^{42}$~cm$^{2}$. Figure \ref{fig:colliders_L} presents impressive progress in the luminosity of colliders -- by more than six orders of magnitude, up to today's record of about $0.5 \cdot 10^{35}$~cm$^{-2}$s$^{-1}$. Note that the luminosity progress goes hand in hand with increase of the energy because the cross-sections of many reactions of interest get smaller with energy and often drop as $\sigma_{reaction} \propto 1/E^2_{cm}$. To get reasonably high numbers of events, one needs to raise the luminosity correspondingly -- as can be seen from Eq.(\ref{eq:intlumi}). For example, for the $WZ$ production in the LHC, with the reaction cross-section of about 6 femtobarn or $6 \cdot 10^{-39}$~cm$^2$, one can expect to see some 1200 of such events over one year of operation (effectively, about $10^7$ s) with peak luminosities $\sim 0.2 \cdot 10^{35}$~cm$^{-2}$s$^{-1}$.

Luminosity of colliders is critically dependent on beam intensities and sizes at the IPs. Colliders usually employ bunched beams of particles with approximately Gaussian distributions, and for $n_b$ bunches containing equal numbers of particles $N_1=N_2=N$ colliding head-on with repetition frequency $f_{\rm rep}$, a basic expression for the luminosity is 
\begin{equation}
    {\cal L} = f_{\rm rep} n_b { N^2 \over 4\pi\sigma_x^{\ast} \sigma_y^{\ast}}\; ,
\label{eq:lumi}
\end{equation}
where $\sigma_x^{\ast}$ and $\sigma_y^{\ast}$ characterize the rms transverse beam sizes in the horizontal and vertical directions at the IPs, respectively. To achieve high luminosity, the  population and number of bunches should be maximized , and either produced as narrow as possible or focused tightly at dedicated colliding locations. Sophisticated detectors usually surround the interaction points in order to collect as much  information as possible about the reactions that originate from collisions of particles.  

In the attempt to understand the ultimate limits of colliders, it should be noted that the great progress of the colliders shown in Figs.~\ref{fig:colliders_E} and \ref{fig:colliders_L} was accompanied by a simultaneous increase of their size, power consumption, complexity and cost. 
Modern colliders employ a number of diverse technologies for power converters and power supplies, ultra-high vacuum systems, particle sources, injection and extraction systems, tunneling, geodesy and alignment, cooling water and cryogenic cooling, beam diagnostics,  accelerator control, personnel safety and machine protection, among other subsystems and equipment. 
Still, when it comes to the facility size, cost and power consumption, the most important factors are the ``core technologies'' required for  accelerating particles to high energies -- normal- and/or superconducting radio-frequency (RF) acceleration systems, and normal- and/or superconducting accelerator magnets -- and ``beam physics techniques'' used to attain the necessary beam qualities such as intensity, beam sizes, and sometimes polarization, including  beam cooling, manipulation and collimation, the production of exotic particles like antiprotons or muons, mitigation of beam instabilities, and countermeasures against beam-size blow up caused by space-charge and beam-beam effects or intra-beam scattering, among other effects.  The energy reach of a collider is mostly defined by its core accelerator technologies, while its luminosity is very much dependent on the sophistication of beam physics techniques \cite{shiltsev2021modern}. 

The energy frontier colliders were and remain costly, often at the brink of financial and political affordability. That poses serious risks, and in the past several projects have been terminated, even after the start of construction. For example, the construction of the 400 GeV c.m.e.~ISABELLE $pp$ collider (briefly renamed CBA) at the Brookhaven National Laboratory in the USA was stopped in 1983 \cite{month2003, crease2005a, crease2005b}; in the early 1990s two other flagship projects were terminated: the 6 TeV c.m.e.~proton-proton complex UNK \cite{yarba1990, kuiper1994} in Protvino, Russia, and the 40 TeV c.m.e.~proton-proton Superconducting Super Collider (SSC) in Texas, USA, in 1993 \cite{wojcicki2009, riordan2015}.
Notwithstanding the above, advances in core accelerator technologies -- including the developments of superconducting magnets for ISABELLE/CBA, UNK and SSC  -- have led to substantial reductions in collider cost per TeV \cite{shiltsev2014}. This progress, together with the growing strength of the high-energy particle physics community, enabled development of frontier machines, such as the currently operational multi-billion dollar LHC.Because no other instrument can replace high-energy colliders in the search for the fundamental laws governing the universe, even larger \$10B-scale future collider projects, need to be motivated and proposed.

\section{Next Few Decades}
\label{future}

The prevailing view of the global HEP community is that the next large particle physics facility should be an $e^+e^-$ collider that functions as a Higgs/ElectroWeak factory. The physics case for such a collider with c.m.e. range (0.25-0.5) TeV and very high luminosity (0.1-1) ab$^{-1}$/yr (hence the name "factory") is quite compelling because it would enable detailed exploration of subtle reactions involving the Higgs/ElectroWeak fields ($H, W, Z$ particles and photons) and shed light on possible deviations from the predictions of the {\it Standard Model} theory of particle physics, see, e.g. \cite{hoddeson1997SMbook, boonekamp2020electroweak}. Several options for each of these types of colliders are under consideration globally, with variable technical readiness. The leading candidates for a Higgs/EW factory are  (1) the $e^+e^-$ Future Circular Collider (FCC-ee) at CERN and the quite similar Circular Electron-Positron Collider (CEPC) in China,  (2) the International Linear Collider (ILC) in Japan, and (3) the Compact LInear Collider (CLIC) at CERN -- see Table \ref{T1}. Additional novel options for compact $e^+e^-$  colliders, such as the Cool Copper Collider (C$^3$), high gradient ($\sim70$ MV/m) superconducting RF linear collider HELEN (High Energy LEptoN collider), ERL-based circular and linear collider schemes, and a Fermilab Site Filler circular $e^+e^-$ collider,  have emerged and are under investigation. For the purpose of this analysis, all Higgs factories can be considered as low-energy machines that can be built based on generally existing technologies and within a reasonable timescale $O$(10-20 years) from the decision to proceed \cite{ITF}. Many beam physics methods and accelerator technologies developed for Higgs factories can be employed in much higher energy machines.

\begin{table}[hbt]
\footnotesize
\centering
\begin{tabular}{| l | c | c | c | c | c | c | c | c |}
\hline
\hline
 Proposal & & Type & $E_{cm}$ & ${\cal L}_{int}$/IP & Yrs. of & Yrs. to 1st & Constr. cost & El. power \\ 
  & & & [TeV] & [ab$^{-1}$/yr] & R\&D & physics & [2021 B\$] & [GW] \\  
\hline
ILC-3 & $e^+e^-$ & L & 3 & 0.61 & 5-10 & 19-24 & 18-30 & $\sim$0.4 \\  
 \hline 
CLIC-3 & $e^+e^-$ & L & 3 & 0.59  & 3-5 & 19-24 & 18-30 & $\sim$0.55 \\  
\hline 
CCC-3 & $e^+e^-$ & L &  3 & 0.6 & 3-5 & 19-24 & 12-18 & $\sim$0.7 \\  
 \hline 
ReLiC-3 & $e^+e^-$ & ERL &  3 & 4.7(9.4)  & 5-10 & $>$25 & 30-50 & $\sim$0.78 \\  
 \hline 
$\mu\mu$Collider$^{1}$-3 & $\mu^+\mu^-$ & C &  3 & 0.23(0.46) & $>$10 & 19-24 & 7-12 & $\sim$0.23 \\  
 \hline 
LWFA-LC-3 & $e^+e^-$ & L &  3 & 1  & $>$10 & $>$25 & 12-80 & $\sim$0.34 \\  
 \hline 
PWFA-LC-3 & $e^+e^-$ & L &  3 & 1  & $>$10 & 19-24 & 12-30 & $\sim$0.23 \\  
 \hline 
SWFA-LC-3 & $e^+e^-$ & L &  3 & 1  & 5-10 & $>$25 & 12-30 & $\sim$0.17 \\  
\hline
 \hline 
Muon Collider$^{1}$ & $\mu^+\mu^-$ & C &  10 & 2(4)  & $>$10 & $>$25 & 12-18 & $\sim$0.3 \\ 
 \hline 
LWFA-LC-15   & $e^+e^-$ & L &  15 & 5  & $>$10 & $>$25 & 18-80 & $\sim$1 \\  
 \hline 
PWFA-LC-15 & $e^+e^-$ & L &  15 & 5  & $>$10 & $>$25 & 18-50 & $\sim$0.62  \\  
 \hline 
SWFA-LC-15 & $e^+e^-$ & L &  15 & 5  & $>$10 & $>$25 & 18-50 & $\sim$0.45  \\  
 \hline 
FNAL $pp$ circ. & $pp$ & C &  24 & 0.35(0.7)  & $>$10 & $>$25 & 18-30 & $\sim$0.4 \\  
\hline
FCC-hh & $pp$ & C &  100 & 3(6) & $>$10 & $>$25 & 30-50 & $\sim$0.56 \\  
\hline 
SPPS& $pp$ & C & 125 & 1.3(2.6)  & $>$10 & $>$25 & 30-50 & $\sim$0.4 \\  
\hline
Collider in Sea& $pp$ & C & 500 & 5  & $>$10 & $>$25 & $>$80 & $>$1 \\
\hline 
\hline 
\end{tabular}
\caption{Main parameters of the multi-TeV lepton collider proposals (3 TeV c.m.e. options) and colliders with 10 TeV or higher parton c.m.e: colliding particles; type of the collider (L for linear, C for circular, ERL for energy recovery linacs); center-of-mass energy (the relevant energies for the hadron colliders are the parton c.m. energy, which is $\sim$ 7 times less than hadron c.m. energy $E_{cm}$ quoted here - see Eq.\ref{eq:eparton}); annual integrated luminosity per interaction point (assuming $10^{7}$s per year effective operating time; for colliders with multiple IPs, the total peak luminosity is given in parenthesis); years of the pre-project R\&D indicate an estimate of the required effort to get to sufficient technical readiness; estimated years to first physics are for technically limited timeline starting at the time of the decision to proceed; total construction cost range in 2021\$ (based on a parametric estimator, including explicit labor, but without escalation and contingency); facility electric power consumption (adapted from the Implementation Task Force report \cite{ITF}).}  
\label{tab:ITF}
\end{table}

At the "energy frontier," the international particle physics community aspires towards a collider with an energy reach of $\sim10$~TeV scale to enable New Physics discoveries (i.e., particles and reactions beyond those described by the Standard Model). The energy of such a collider should significantly exceed the 14 TeV c.m.e. of the LHC, which can be provided either by a $\sim100$~TeV hadron ($pp$) collider or a $\geq10$ TeV lepton ($e^+e^-$ or muon) collider. Here it should be noted, that in very high energy collisions, hadrons manifest themselves as composites of quarks and gluons, whose total energy is distributed among these constituents. Therefore, the highest accessible c.m.e. $E^*_{cm}$ of individual parton-to-parton collisions is significantly lower than the nominal (proton-proton) $E_{cm}=2E$, and, e.g., for many reactions it can be assumed that \cite{muonsmashers2022}:
\begin{equation}
E^*_{cm} \simeq (1/7-1/10) \times E_{cm} =(1/7-1/10) \times 2E.
\label{eq:eparton}
\end{equation}
Several $\sim10$~TeV c.m.e. scale collider options are under active discussion at present -- see Table \ref{tab:ITF} -- including two $pp$ colliders, the FCC-hh at CERN and SPPC in China; 3 TeV to 14 TeV muon colliders; as well as novel $e^+e^-$ collider schemes based on plasma wakefield acceleration. 

 In the course of the recent {\it Snowmass'21} US community strategic planning exercise, the Implementation Task Force (ITF) \cite{ITF} of a dozen internationally renowned accelerator experts was convened and charged with developing metrics and processes to facilitate comparisons between projects. Essentially all ($>$ 30) collider concepts presently considered viable have been evaluated by the ITF using parametric estimators to compare physics reach (impact), beam parameters, size, complexity, power, environmental concerns,  technical risk, technical readiness, validation and R\&D required, cost and schedule -- see Table \ref{tab:ITF}. The significant uncertainty in these values was addressed by giving a range where appropriate. Notably, the ITF choose to use the proponent-provided luminosity and power consumption values. 
Relevant measure of the maturity of a proposal is the estimate of how much R\&D time is required before a proposal could be considered for a project start (so called "Critical Decision 0" in the US scientific infrastructure project approval system). The time to first physics in a technically limited schedule includes the pre-project R\&D, design, construction and commissioning of the facility, and is most useful to compare the scientific relevance of the proposals over the timeline of interest. 

The total project cost follows the US project accounting methods, but without taking into account the inflation escalation and (usually required) contingency. The ITF used various parametric cost models, also taking into account the estimates provided by the proponents, and -- for reference -- known costs of existing installations, as well as reasonably expected costs for novel equipment. For future technologies, the pre-project cost reduction R\&D may further lower the ITF cost estimate ranges.

As for any large scientific research facility, it is not only the cost that is of importance, but also the number of experts needed for the design, construction and commissioning of the future colliders and the environmental impact, e.g., the electrical power consumption. Therefore, it is of very practical interest for the particle physics community to assess the limits of the ultimate colliders in a quantitative manner. 

\section{Limits of Colliders}
\label{limits}

A discussion of the limits of future colliders starts with an introduction to the issue: definitions of the units and general considerations regarding energy, luminosity, and social cost of the ultimate machines. It is followed by a more detailed look into specific limitations of circular $pp, ee$ and $\mu\mu$ colliders; linear and plasma-based $ee, \gamma \gamma, \mu\mu$ colliders; and some exotic schemes, such as the crystal muon colliders. The social-cost considerations (power consumption, financial costs, carbon footprint, availability of experts and time to construct) are most defined for the machines based on extensions of the existing core accelerator technologies (RF and magnets) and less so for the emerging or exotic technologies (ERLs, plasma WFA, crystals, etc). 

Three of the most important aspects of the evaluation are feasibility of the c.m. energy $E_{cm}$, feasibility of the collider luminosity $\cal L$, and feasibility of the facility cost $C$.  For each machine type (technology),  current state-of-the-art machines are examined -- see, e.g., Ref.\cite{shiltsev2021modern} for more details --  and several (1,2,...) orders of magnitude steps in energy made to see how that would affect the luminosity and cost. 

The unit of the c.m. energy  $E_{cm}$ used is 1 PeV = 1000 TeV. The units of $\cal L$ are ab$^{-1}$/yr, i.e., equal to, e.g., $10^{35}$ cm$^{-2}$s$^{-1}$ over $10^7$ sec/yr. For reference, the LHC will deliver 0.3 ab$^{-1}$/yr after its high luminosity upgrade. Due to the spread of expectations for the machine availability and annual operation time, there might be a factor of $\sim$2 uncertainty in peak luminosity for any ab$^{-1}$/yr value. The units of  electric power consumption are TWh/yr. For reference, the CERN power consumption averages about $P_s$=200MW and 1.1-1.3 TWh/yr while operating the LHC. The total facility electric power includes not only the collider and its injectors, but also detectors, infrastructure, lighting, etc. In addition, accelerator systems needed to maintain and accelerate beams (RF, magnets, etc) have their own inefficiencies, and as result, for all collider types the facility electric power is significantly larger than the power that goes into the beams.  The cost is estimated in "LHC-Units" (LHCU) -- the cost of the LHC construction (at present day prices,  LHCU$\simeq$\$10B). An analysis similar to that of the ITF is the most reliable (see above Sec.\ref{future}). With certain reservations and caveats, an approximate phenomenological $\alpha \beta \gamma$ collider cost model \cite{shiltsev2014} is appropriate: 
\begin{equation}
C_{c} \approx \alpha \cdot L_t^{p_1}
 + \beta \cdot E_{cm}^{p_2} + \gamma \cdot P_s^{p_3}
\label{ModelCost}    
\end{equation} 
where the cost is understood as a total project cost (of an all-new facility without previous investments, taking into account labour cost, escalation due to inflation, contingency, R\&D, management, etc.) that scales with just three facility-specific parameters — the length of the tunnels $L_t$, the center-of-mass or beam energy $E_{cm}$, and the total required site power $P_s$. The second term reflects the cost of accelerator components (magnets, RF, etc. and associated auxiliary subsystems); it depends very much on technology and often dominates the total cost.  Comparison with the cost of recently built large accelerators and the ITF cost estimates indicates that the model estimates are good to within a factor of 2 if the exponents are rounded up to $p_1=p_2=p_3=1/2$, and the coefficients are $\alpha \approx$ 0.1LHCU/$\sqrt{10 \rm{km}}$, $\gamma \approx$ 0.3LHCU/$\sqrt{\rm{TWh/yr}}$ and the accelerator technology dependent coefficient $\beta_{\rm MAG} \approx$6LHCU/$\sqrt{\rm{PeV}}$ for high-field magnets and $\beta_{\rm RF} \approx$30LHCU/$\sqrt{\rm{PeV}}$ for RF accelerating structures \cite{shiltsev2014, ITF}. The $\alpha \beta \gamma$-model should be used with caution as it still needs to be properly extended to advanced technologies (plasma WFA, lasers, crystals, etc). 

\subsection{General Limitations}

The most obvious limit to consider is the size of the collider. Indeed, as Eqs.\ref{eq:energy_l} and \ref{eq:energy_c} indicate, the larger the length of a linac or circumference of a ring, the higher beam energies $E$ can be envisaged. For example, if the available site length is limited to $L_t \simeq 100$ km, then two linacs of 50 km each  could  allow the energy to reach up to $E_{cm} \simeq$0.01 PeV with the current state of the art normal-conducting RF cavities with $G=0.1$ GeV/m and up to $E_{cm} \simeq$0.2-0.5 PeV with the potentially achievable average accelerating gradient of $G=2-5$ GeV/m in plasma-wakefield structures. In comparison, a 100 km-long circular tunnel ($\rho$=16 km radius) allows a $\sim$0.1 PeV collider based on the 16T Nb$_3$Sn SC bending magnets or a 0.25 PeV collider with $\sim$40T high-temperature superconducting (HTS) magnets. Of course, larger circumference tunnels could fit proportionally higher c.m. energy machines. 

Note that not all kinds of particles can be accelerated in high-energy circular colliders due to prompt synchrotron radiation (SR) that results in the energy loss per turn of 
\cite{Sands:1970ye}:
\begin{equation}
\Delta E_{SR} = \frac{1}{3 
\epsilon_{0}} \frac {e^2 
\beta^{3} \gamma^4}{\rho} \, , 
\label{SR}
\end{equation}
which increases with the fourth power of energy $E=\gamma mc^2$ and scales with the inverse of the bending radius (here, $\epsilon_0$ is the permittivity of the vacuum and $\beta=\sqrt{1-1/\gamma^2}$).  
At the limit of practicality, the SR loss per turn should be at least less than the total beam energy $\Delta E_{SR} \leq E$, which defines the c.m. energy limit for circular colliders as:
\begin{equation}
\label{eqRingLimit}
    E_{cm}{\rm{[PeV]}} \leq 0.001 \cdot (m/m_e)^{4/3} (\rho/10 {\rm{[km]}})^{1/3} \, \, \, ,
\end{equation}
assuming $\rho \sim 10$ km, that is $\sim$1 TeV for electrons, $\sim$1.2 PeV for muons ($m_\mu\approx$210$m_e$) and $\sim$25 PeV for protons  ($m_p\approx$2000$m_e$). Beyond these energies, sheer energy economy will demand that colliders be linear (thus, needing no bending magnets). 

Survival of the particles in very long accelerators may set another energy limit. Indeed,for example, a 0.5 PeV linear collider based on individual 5 GeV plasma-wakefield accelerating stages requires $M=10^5$ of them. For a beam of particles to propagate through such a chain without losing too much intensity (and power), the stage-to-stage transfer efficiency must be much better than $\eta_{stage} \ge 1-1/M=0.99999$ -- an extremely difficult challenge. Also, if the particles are unstable, they may decay before the end of the acceleration process. To guarantee delivery to the collision point, the minimum accelerator gradient must significantly exceed $G \gg mc/\tau_0$ -- where $\tau_0$ is the proper decay time - that is, e.g., 0.3 MeV/m for muons (relatively easy to achieve even with present day technologies) and 0.3 GeV/m for tau-leptons (quite a challenge even for the most optimistic currently envisioned advanced acceleration schemes) \cite{shiltsev2012ufn}. 

Performance (luminosity) reach of the ultimate colliders can be limited by many factors and effects -- particle production, beamstrahlung, synchrotron-radiation power per meter, IR radiation damage, neutrino-radiation dose, beam instabilities, jitter/emittance growth, etc -- which are machine specific and will be considered below. However, the most fundamental is the limit on the total beam power $P_b=2 \times f_0 n_b N \gamma mc^2$ (the factor of 2 accounts for two colliding beams). Indeed, the luminosity equation (\ref{eq:lumi}) can be re-written as:  
\begin{equation}
\label{eq2}
    {\cal L} = {1 \over 16 \pi f_{rep} n_b \varepsilon \beta^* m c^2}\cdot {P_b^2 \over E} \propto {P_b^2 \over E} \, ,
\end{equation}
where $\sigma_x^*\sigma_y^*=\varepsilon_n \beta^* / \gamma$ has been substituted with so-called "normalized beam emittance" $\varepsilon_n$ and the so-called "beta-function at IP" $\beta^*$, which is generally not explicitly dependent on energy - see \cite{shiltsev2021modern}. Particle accelerators in their essence are transformers of wall-plug site power $P_s$ into high-energy beam power $P_b=\eta P_s$ with much less than 100\% efficiency (in the best-case scenario $\eta \sim 0.1-0.3$). It is hard to know precisely where the ever-changing societal limits on the power consumption of large accelerators will be in the future, but they will surely include "carbon footprint" considerations and the environmental impact of future accelerators' construction and operation. For reference, with the present world-average power consumption rates, 1,000,000 people require $\sim$3TWh/yr, which is three time larger than the CERN annual site energy usage.  Wherever the limit is, Eq.(\ref{eq2}) points out that the luminosity will decrease with energy at least as $L \propto 1/ E$. Such dependence on energy is markedly different from the traditional HEP demand for the luminosity to follow the point-like annihilation cross-section scaling, $L \propto E^2$; from current knowledge, other factors  $f_{rep}, \, n_b, \, \varepsilon_n, \, \beta^*, \, \eta$ could be of only limited help in avoiding performance degradation in the quest for two to three orders of magnitude higher energies. 

Of course, there will also be societal limits on the collider's total cost $C_c$. While this depends on the technology (core accelerator technology, civil construction technology, electric-power production, delivery and distribution technology, etc.), the probability of approval and realization for a technically feasible future collider facility typically decreases with cost increase beyond what is "reasonably acceptable", perhaps as $\propto 1/C_c^\kappa$. As a guide, such a decrease, with the exponent $\kappa\approx 2-3$, is characteristic for the price distributions of real-estate sales. Also note: i) the costs of civil construction and power systems are mostly driven by the larger economy and are not that dependent on the collider type and accelerator R\&D advances; ii) if an injector complex is already available, up to 1/3 of the total cost could be saved, resulting in potential increase of a factor of 2 in the energy reach - see Eq.(\ref{ModelCost}); iii) the collider cost is usually a relatively weak function of luminosity, provided new technologies are not required (the latest example is the HL-LHC \$1B project that will increase luminosity of the $O$(\$10B) LHC by a factor of 5); iv) future machines are best designed with high $E$ and relatively low initial $\cal L$ in anticipation of eventual performance upgrades (for example, in the past, CESR and the Tevatron witnessed $\cal L$ increases $O$(100), LHC by a factor $\ge$10, etc.); v) the total cost $C_c$ is moderately weakly dependent on the tunnel length/circumference $L_t$, but it is critically dependent on $E_{cm}$ and the choice of the acceleration technology. 

The construction time of large accelerator projects to date is usually between 5 and 11 years and approximately scales as $T \propto \sqrt{C_c}$. It is often limited by the peak annual spending rate, typically in the range \$0.2 to \$0.5 B/yr (compare to the world’s global HEP budget $\sim$\$4B), which in turn depends on the number of available technical experts. So far, the period of technical commissioning of colliders, often defined as “one particle reaches design energy”, was $O$(1) yr – and is shorter for known technologies and longer for new ones and for larger numbers of accelerator elements. Progress towards the design (or ultimate) luminosity is dependent on the machine's “complexity" \cite{shiltsev2011acomplexity}, and can take as long as $\sim$9 yrs \cite{ITF}.

Taking all the above into account, various types of future colliders are analysed below and their potential energy and luminosity reach assessed -- maximum $E_{cm}$ and peak $\cal L$ -- under the assumption of the societal limits on the site power consumption and cost: 
\begin{equation}
\label{limits}
    P_s \, \leq 3 \, {\rm TWh/yr} \, \ , {\rm and} \, \, \, C_c \, \leq 3 \, {\rm LHCU} .
\end{equation}

\begin{figure}[htbp]
\centering
\includegraphics[width=0.99\linewidth]{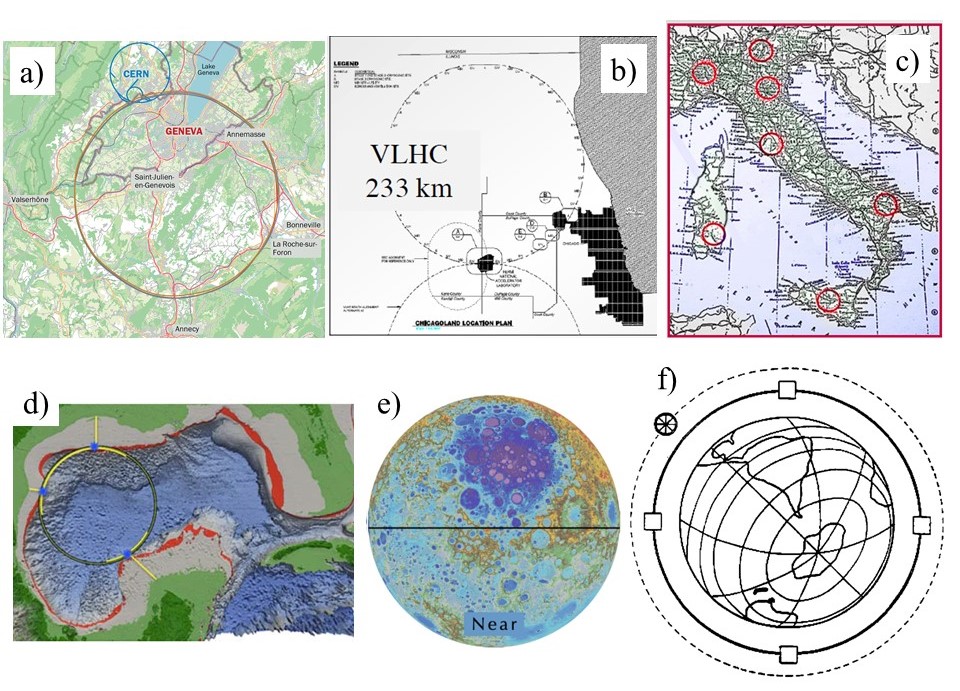}
\caption{Very large hadron collider proposals (not in scale): a) FCChh (91 km circumference, 100 TeV), b) VLHC (233 km, 175 TeV); c) Eloisatron (300km, 200 TeV); d) ``Collider in the Sea" (1,900 km, 500 TeV), e) collider on the Moon (11,000 km, 14 PeV), f) Enrico Fermi's accelerator encircling our Earth ("Global-tron", 40,000km, 2.9 PeV)} 
\label{fig:hh}
\end{figure}

\subsection{Circular $e^+e^-$ colliders}
As mentioned above, the synchrotron radiation of light leptons $e^+, e^-$ limits the energy reach of such colliders to $E_{cm} \le 1$ TeV, which is far below even the energy reach of the LHC, to say nothing the aspiration to reach to PeV energies. High luminosity could be a potential rationale for an interest in these types of colliders, but it is limited by synchrotron-radiation power losses $P_{SR}=2 f_{\rm revolution} e n_b N \cdot \Delta E_{SR}$ and very quickly drops with energy as:  
\begin{equation}
 {\cal L}_{ee \, \rm{cir}} = {\cal F}_{ee}
 \frac {P_{\rm SR} \rho}{ \gamma^3} \, , 
\label{eq:lumi3}
\end{equation}
The factor ${\cal F}_{ee}$ above accounts for the IP vertical focusing parameters and a dimensionless {\it beam-beam parameter} that reflect the severity of the electromagnetic disruption of one beam after collision with another - the exact expression is given elsewhere \cite{shiltsev2021modern}.  Of importance for this discussion is that   ${\cal F}_{ee}$ is weakly dependent on the beam energy, and the maximum practical luminosity of $e^+e^-$ circular colliders scales as $1/E^{3-3.5}$. These facilities naturally call for larger radius $\rho$ and circumference $O$(100 km) -- see Eq.(\ref{eq:lumi3}) -- and are considered quite promising tools at low energies, e.g., as high-luminosity Higgs/ElectroWeak factories with typical $E_{cm} \simeq 0.25$ TeV, but even these have an energy demand of some (1.5-2) TWh/yr and cost $\sim$(1.5-2) LHCU. Significant energy savings are possible by using  RF energy-recovery (ERLs), but that expands the c.m.e. reach of circular $e^+e^-$ colliders to only $E_{cm} \sim $ 0.5 TeV. 

\begin{figure}[!htb]
   \centering
   \includegraphics*[width=.90\columnwidth]{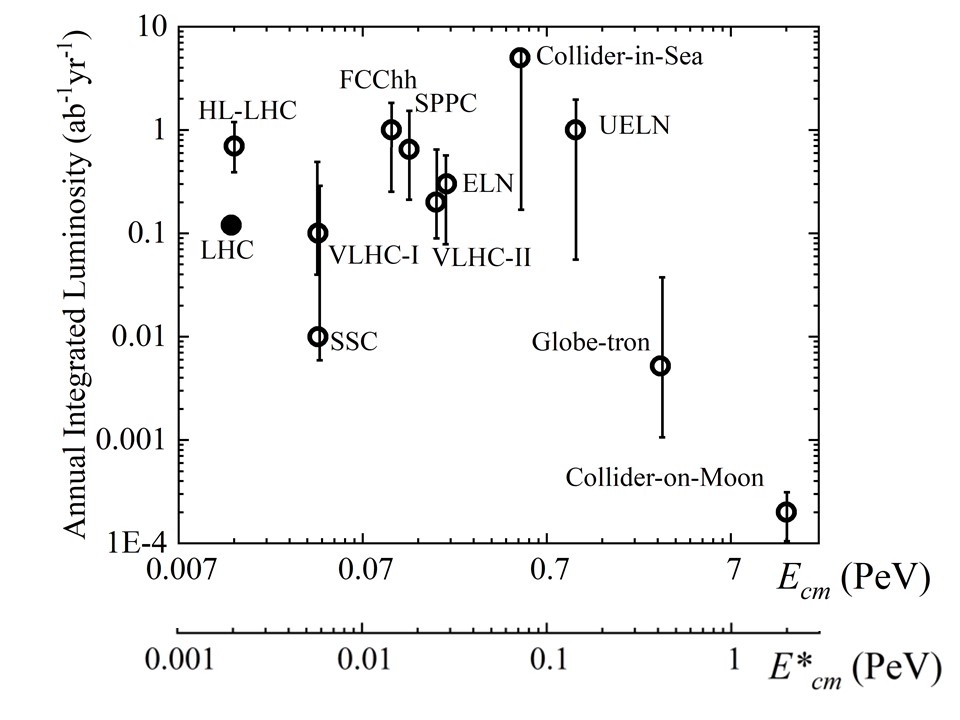}
   \caption{Estimates of the annual integrated luminosity for very high energy circular hadron colliders vs $E_{cm}$. The second horizontal axis is for the approximate equivalent parton center-of-mass energy $E^{*}_{cm} \approx E_{cm}/7$. }
   \label{figPPcirc}
\end{figure}

\subsection{Circular $pp$ colliders}

Being significantly less limited by the synchrotron radiation losses -- see defined Eq.(\ref{eqRingLimit}), protons can be accelerated in circular machines to multi-PeV energies, and, according to Eq.(\ref{eq:energy_c}), the limit is fully determined by the maximum field $B$ of the bending magnets and the tunnel circumference $L_t \simeq 2 \pi \rho$. Fig.\ref{fig:hh} presents several $pp$ collider proposals aimed for higher and higher energies which are based on increasing either $B$ or $L_t$ or both. Most appropriate magnet technologies currently assume limits on the maximum bending field: about 2 T for normal-conducting magnets (usually, room temperature copper conductor and steel yoke), some 8 T for NbTi SC technology, up to 16 T for Nb$_3$Sn SC technology \cite{zlobin2019nb3sn}, and 20 T to (max) $\sim$40 T for high-temperature superconductor (HTS) technologies (e.g, based on rare earth oxides like ReBCO, or iron-based superconductors). 

There is significant knowledge in the physics community on how to design, build and operate circular $pp$ colliders -- e.g., experience with the Tevatron $p \bar p$ collider ($E_{cm}$=0.002 PeV, $B=$4.5T, 6.3 km circumference) \cite{lebedevshiltsev2014tevatron} and 0.014 PeV LHC (8T, 27km) \cite{evanslhc}.  Also, there are designs and/or parameter sets available for the Superconducting Super Collider (SSC, 0.04 PeV, 6.6T, 87km), Future Circular $pp$ Collider (FCC-hh, 0.1 PeV, 16T, 91km) \cite{fcchh}, Super proton-proton Collider (SppC, 0.075-0.125 PeV, 12-24T, 100km) \cite{cepc}, Very Large Hadron Collider (VLHC, 0.175 PeV, 12T, 233km) \cite{accel:vlhc}, the Eloisatron (0.3 PeV, 10T, 300km) \cite{barletta1996eloisatron}, and "Collider-in-the-Sea" (in the Gulf of Mexico, 0.5 PeV, 3.2T, 1900km) \cite{mcintyre2017sea500Tev}. Going to the extreme, Enrico Fermi had thought of an accelerator encircling the Earth which could reach about 3 PeV c.m.e. with inexpensive normal-conducting magnets \cite{cronin2004fermi}, and, more recently, a circular collider on the Moon was discussed (CCM, 14 PeV, 20T, 11,000 km) \cite{zimmermann2022moon}. 

The most stringent limitations come from large size (related to the magnetic field $B$ technological limit) and very high power-consumption requirements, resulting in high cost.
Already the 100-km machines like the FCChh and SPPC could be approaching an energy need of 3 TWh/yr and be over the 3 LHCU cost limits given by Eq.\ref{limits}. Of course, even that is small in comparison with the lunar CCM cost (about 20-40 LHCU just for the SC magnets) and the energy needs, which are (2-5)$\cdot 10^{4}$ TWh/yr ($O$(30\%) of the world's current production). 

Even more serious are limitations on the maximum attainable luminosity ${\cal L}_{pp}$. With the increase of beam energy, limiting detrimental effects include beams disruption due to opposite bunch EM forces experienced at each IP (beam-beam effects) and coherent beam instabilities induced by the beams' own EM interaction with induced image charges, currents and wakefields (especially dangerous in large-circumference high-intensity machines). 
Unavoidable will be fast beam burn-off -- destruction due to inelastic interactions of high-energy protons as the result of repetitive collisions -- leading to shorter and shorter beam lifetime:
\begin{equation}
\tau_{pp} = \frac {n_b N} {{\cal L}_{pp} \sigma_{tot}} \, . 
\label{eq:burnoff}
\end{equation}
The total $pp$ cross-section grows slowly from $\sim$100 mbarn to $\sim$300 mbarn with an increase of $E_{cm}$ from 0.001 PeV to 1 PeV. The burn-off at very high energies results in several undesired effects: first, a short beam lifetime $\tau_{pp}$ of about an hour or even minutes, which requires the frequent injection and acceleration of new bunches of particles. Injection and acceleration in a chain of SC magnet-based boosters is a lengthy process and, therefore, a smaller fraction of the operation time is left for collisions and the entire accelerator complex efficiency drops. Secondly, the particle detectors get flooded with products of the inelastic interactions -- the so-called {\it pile-up} effect makes it extremely difficult to disentangle the huge number of tracks originating from approximately 1000 or more $pp$ reactions per bunch collision with luminosity $O$(10$^{35}$ cm$^{-2}$s$^{-1}$). Thirdly, growing problems are also anticipated with radiation protection of the detectors and collider elements and collimation of beams  with higher energy density. 

For $pp$ colliders with $E_{cm}$ above (0.1-0.2)PeV, synchrotron radiation will essentially limit the maximum attainable luminosity in very much the same fashion as for $e^+e^-$ colliders -- see Eq.(\ref{eq:lumi3}) -- because of either the limited RF power available to replenish the SR losses $P_{SR}$ or due to challenges related to the cooling of the SC magnets, where the SR photons must be intercepted internally and significant heat load due to these photons needs to be extracted at cryogenic temperatures. 

Individual machine designs may vary in optimization approaches toward the highest luminosity; Fig.\ref{figPPcirc}  presents estimates of performance of circular $pp$ colliders vs c.m. energy up to $E_{cm}=$14 PeV (equivalent to parton c.m. energy $E^*_{cm} \simeq 2$ PeV, according to Eq.(\ref{eq:eparton})). Even with the logarithmically large uncertainties (indicated by the error bars) in scenarios limited by electric power, very high energy colliders will by necessity have low luminosity.

\subsection{Circular $\mu\mu$ colliders}
Colliding muons would have two key advantages: i) compared to protons, the same size machine would allow effectively a factor of 7-10 higher energy reach due to the point-like nature of the muons -- see Eq.(\ref{eq:eparton}); and ii) according to Eq.(\ref{SR}), the synchrotron radiation of muons is $\sim (m_\mu/m_e)^4=2$ billion times weaker than that of electrons and positrons, and power- and cost-effective acceleration in rings is possible to about a fraction of a PeV -- see  Eq.(\ref{eqRingLimit}). Therefore, the highest energy circular muon colliders are predicted to be more compact, more power-efficient and significantly less expensive than the equivalent energy-frontier hadron or $e^+e^-$ machines \cite{long2021muonnature}.  

These advantages come along with difficulties due to the short lifetime  of  the  muon,  $\gamma \tau_0$ where $\tau_0$=2.2$\mu$s. For example, a 0.1 PeV $\mu^-$-meson has on average a lifetime of one second, decaying into an electron (or positron in the case of $\mu^+$ decay) and two neutrinos, each carrying a significant fraction of the initial muon momentum.  It is widely believed that the time before the decay is more than sufficient to allow fast acceleration of muons to high energy, followed by a storage for some 300$B$ turns in a ring  with an average bending magnet field $B$ (in units of Tesla) where $\mu^-$ and $\mu^+$ particles will collide with each other \cite{sessler1998muoncollphystoday}. 

\begin{figure}[htbp]
\centering
\includegraphics[width=0.99\linewidth]{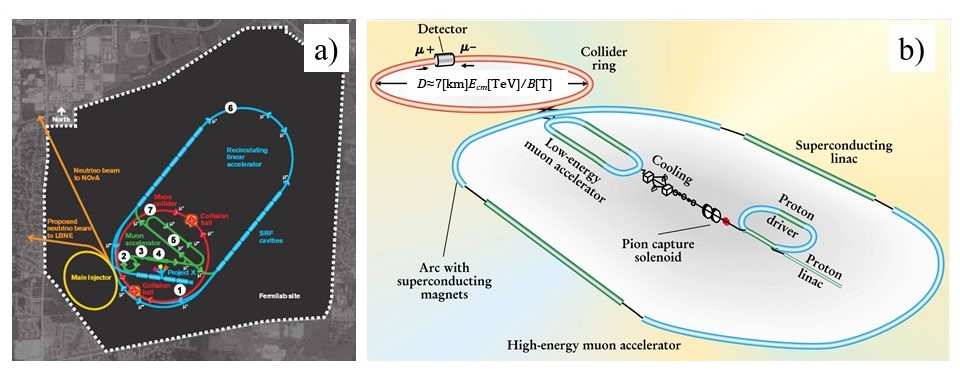}
\caption{Schematics of high-energy circular muon colliders: a) on the FNAL site, b) a general scheme (adapted from \cite{sessler1998muoncollphystoday}).} 
\label{fig:mumu}
\end{figure}

As schematically shown in Fig.\ref{fig:mumu}, a $\mu^+ \mu^-$ collider will not look much different from the $pp$ collider rings -- it will consist of accelerating RF cavities and high-field (SC) magnets, the latter determining the size of the facility for a given $E_{cm}$. What will be different is a somewhat more complicated system of production of the muons in the reactions resulting from multi-GeV protons hitting stationary targets, collection of these muons, muon beam {\it cooling} (significant reduction of the muon beam sizes and internal velocity spreads), and rapid acceleration to the energy of the collider \cite{palmer2014muonrast}. 

There are parameter sets available for 1.5, 3, 6, 10, and 14 TeV circular $\mu\mu$ colliders which indicate their superior (w.r.t. other collider types) power efficiency in terms of ab$^{-1}$/TWh \cite{shiltsev2021modern}. Projecting their site power requirements and costs, ``the feasibility limits" of 3TWh/yr and 3 LHCU -- see Eq.(\ref{limits}) -- will take place at $E_{cm}$=(0.03-0.05) PeV. 

\begin{figure}[!htb]
   \centering
   \includegraphics*[width=.99\columnwidth]{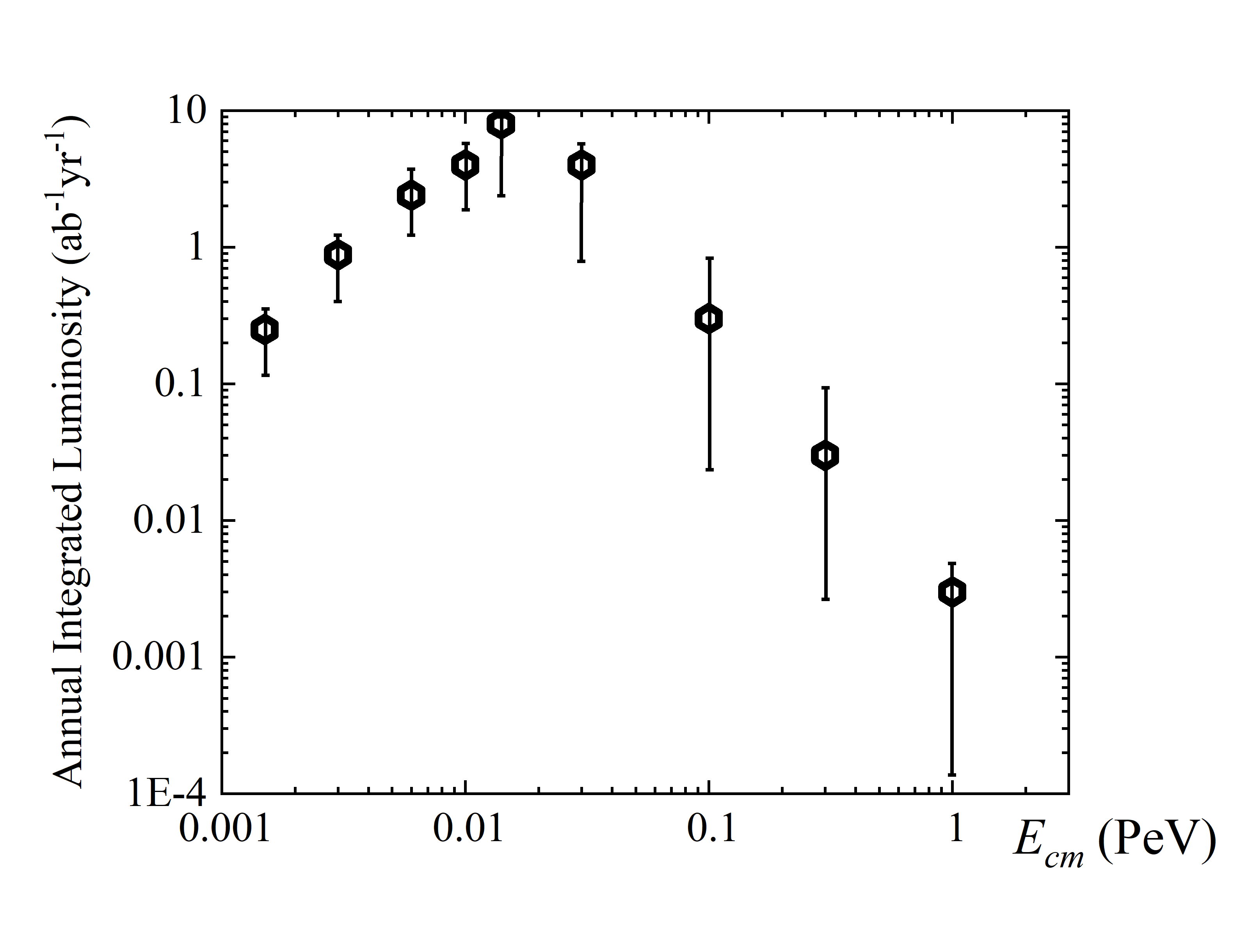}
   \caption{Estimates of the annual integrated luminosity for very high-energy circular muon colliders.}
   \label{fig3}
\end{figure}

The average luminosity of a muon collider is equal to:  
\begin{equation}
 {\cal L}_{\mu\mu} = f_{rep} \gamma \frac{c\tau_0}{4 \pi \rho} 
\frac{ n_{\rm b} N^2 }{ 4\pi {\sigma^*_x \sigma^*_y}} = {\cal F}_{\mu \mu} B P_b \gamma \, ,
\label{eq:lumi5}
\end{equation}  
where $f_{rep}$ is the rate of the facility acceleration cycles. The luminosity can be seen to scale with $B$ and with the total beam power $P_{b}=f_{rep} en_bNE_{cm}$. Exact expression for the factor ${\cal F}_{\mu\mu}$ can be found in, e.g., \cite{palmer2014muonrast}. 
The above Eq.(\ref{eq:lumi5}) indicates an obvious incentive to have the highest bending magnetic field $B$ and the luminosity increase with energy $ {\cal L}_{\mu \mu} \propto \gamma$, in the case of other limiting parameters fixed. 

Unfortunately, above about 0.01 PeV, the intense neutrino flux originating from the muons decaying in the collider poses the challenge of minimizing the environmental impact. The collider complex is usually located underground, and when the produced neutrinos emerge at the surface, a small fraction interacts with the rock (and other material) and produces an ionizing radiation dose that quickly grows with energy $D_\nu \propto f_{rep} n_b N E^3$. The impact of this neutrino-induced radiation can be mitigated, for example, by continually adjusting the orbits of the beams to spread them out on a wider area, by deeper collider tunnels  or by a further reduction of the emittance of the muon beam so that the required luminosity could be obtained using a substantially smaller number of muons. It is believed that the neutrino flux dilution factor $\Phi$ could be as high as 10-100 and the ultimate luminosity will depend on it as: 
\begin{equation}
\label{eq3}
    {\cal L}_{\mu\mu} \propto {D_\nu \Phi \over E_{cm}^2} \, .
\end{equation}
Additional uncertainty at high energies will be limited capabilities to operate SC magnets with significant deposition of the beam power inside them -- muons decay into high-energy electrons, which will be quickly bent by the strong magnetic field into the vacuum chamber/absorber walls, radiating SR on their way.

Therefore, the resulting luminosity projections for muon colliders indicate a promising increase up to $E_{cm} \sim$0.02 PeV followed by fast decline, approximately as shown in Fig.\ref{fig3}.  

\subsection{Traditional, advanced and exotic linear $ee$ or $\mu \mu$ colliders}

\begin{figure}[htbp]
\centering
\includegraphics[width=0.99\linewidth]{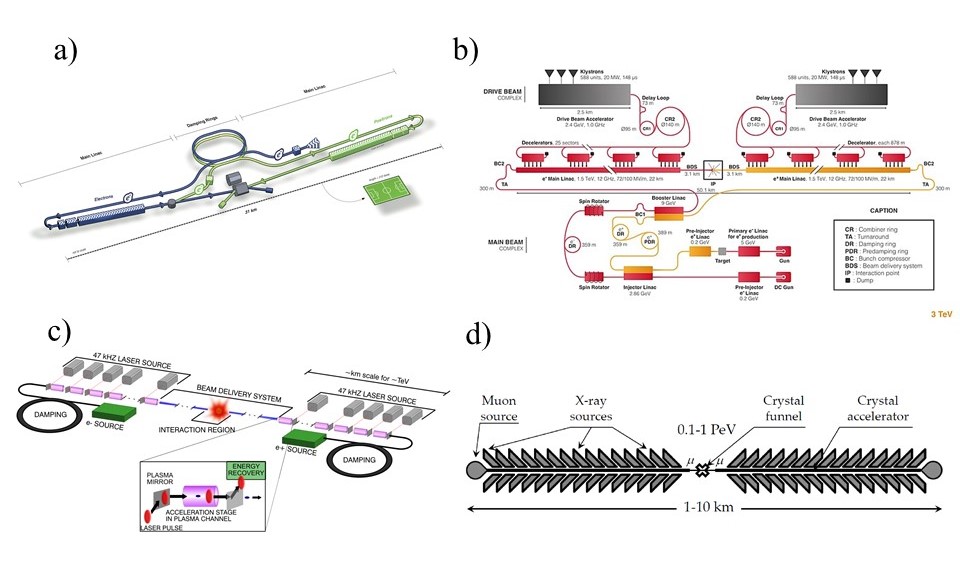}
\caption{Very high energy linear lepton collider proposals (not to scale): a) 1 TeV c.m.e. $e^+e^-$ ILC (31 km long), b) 3 TeV c.m.e. $e^+e^-$ CLIC (50 km); c) plasma wakefield linear $e^+e^-$ collider (length depends on energy, e.g., $\sim$20 km for 30 TeV c.m.e.); d) linear crystal wakefield $\mu^+\mu^-$ collider.} \label{fig:ll}
\end{figure}

Acceleration in linear systems (without bending magnets) allows, in principle, the avoidance of the energy limits of Eq.(\ref{eqRingLimit}) due to the absence of synchrotron radiation (the power of a particle's radiation in a longitudinal field is $\gamma^2$ times smaller than in an equivalent transverse field). A huge disadvantage of linear colliders (LCs) is that beams are used (collide) only once and then are spent in beam dumps -- that leads to intrinsic power inefficiency and, as shown below, low luminosities. The energy limit will be set by the available length of the tunnel and the average accelerating gradient. In traditional RF accelerating structures, the latter is limited to $G_{RF} \sim$0.2 GV/m by the structure damage due to discharges. As a result, even the most optimized traditional LC designs, like the ILC\cite{michizono2019ilcmature} and CLIC\cite{stapnes2019clicnature}, become quite long (30-50 km, see Figs.\ref{fig:ll}a and b) and get to the limits on cost and power limits (3 LHCU and 3 TWh/yr) already at 0.001-0.003 PeV.  

Ionized plasmas can sustain electron plasma density waves with accelerating electric field gradients up to: 
\begin{equation}
G_p=m_e \omega_p c/e\approx 0.1\, {\rm [TV/m]}\, \sqrt{n_0{\rm [10^{18} \, {\rm cm}^{-3}]}},
\label{plasmagradient}
\end{equation} 
where $n_0$ denotes the ambient electron number density and $\omega_{\rm p}=\sqrt{e^2 n_0/(m_e \varepsilon_{0})}$ is the electron plasma frequency \cite{tajima1979laser}. Such gradients can be effectively excited by either powerful external pulses of laser light or electron bunches if they are shorter than the plasma wavelength $\lambda_{\rm p}=c/\omega_{\rm p} \approx1$~mm$\times \sqrt{10^{15}\, {\rm cm^{-3}}/n_0 }$, or by longer beams of protons if their charge density is modulated with the period of $\lambda_p$ \cite{gonsalves2019petawatt8Gev, litos20169gev, adli2018awakenature}. Whether plasma acceleration will be suitable for a very high energy collider application is yet to be seen, given the necessity of very high efficiency staging and phase-locking acceleration in multiple plasma chambers \cite{leemans2009lpwaphystoday, schroeder2010lpwacolliderprab}. Also, at the present early stage of development of this advanced plasma-wakefield technology, the cost of such a collider would be extremely high and a potential for the several orders of magnitude improvement in the cost efficiency still needs to be demonstrated. It is clear, though, that any type of linear collider will be power-hungry. Indeed, its luminosity scales as:
\begin{equation}
\label{eq:lumiLC}
 {\cal L}_{\rm lin} = {\cal F}_{\rm lin} {N_\gamma \over \sigma_y^*} {P_{b} \over E_{cm}} \, , 
\end{equation}
which decreases at higher energies if the  total beam power is limited. Other factors in the equation above are limited too, such as the beam sizes at the IP $\sigma^*_{x,y}$ (strongly dependent on the jitter of the collider elements and sophistication of the final-focus system) and $N_{\gamma}\approx 2 \alpha r_0 N/\sigma_{x}^{\ast}$ -- the number of beamstrahlung photons emitted per $e^{\pm}$ ($\alpha \approx 1/137$ denotes the fine-structure constant). The latter characterizes the 
energy radiated due to the electromagnetic field of  one bunch acting on the particles of the other  ({\it beamstrahlung}) and the corresponding c.m. energy spread that should be controlled to be  $\ll E_{cm}$ -- further details and the exact expression for ${\cal F}_{\rm lin}$ can be found in, e.g., \cite{shiltsev2021modern}. 

Most technologically feasible are LCs colliding electrons with electrons, but the particle physics reach of high-energy $e^-e^-$ collision (variety of possible reactions and their cross-sections) is significantly less inspiring than that of the $e^+e^-$ colliders -- see, e.g., \cite{barger2018collphysbook}. In order to avoid the c.m.e. spread induced by the beamstrahlung, which at high energies $E_{cm} \ge 3$TeV and luminosities approaches 100\%, conversion of electrons into photons -- via inverse Compton scattering on the high-brightness laser beam right before the IP -- was proposed \cite{ginzburg1983colliding}. The resulting $\gamma \gamma$ collisions would have kinematic advantages for some HEP reactions, though still with significant c.m.e. spread. Proton linear colliders have never been seriously considered because of the factor of 7-10 disadvantage in the effective c.m. energy reach w.r.t. leptons - see Eq.(\ref{eq:eparton}). Until recently, linear muon colliders were not discussed either, due to obvious difficulties with muon production and collection. An interesting opportunity of wakefield acceleration of muons in structured solid media, e.g., carbon nanotubes(CNT) or crystals with the charge carrier density $n_0\sim$10$^{20-22}$ cm$^{-3}$, was proposed in \cite{tajima1979laser}. It promises extreme accelerating gradients 1-10 TV/m, continuous focusing and simultaneous acceleration (no cells, one long channel, particles get strongly cooled via the betatron radiation while channeling between the crystal planes or inside individual CNT channels). A corresponding linear crystal muon collider \cite{chen1997crystal, shiltsev2012ufn} would be compact in size ($\sim$10 km for 1 PeV) -- see, Fig.\ref{fig:mumu} d) - and, therefore, have the promise of low(er) cost. The luminosity of such an exotic LC would still be very low -- $O$(0.1 ab$^{-1}$/yr) at best -- for the same reasons as for any linear collider. 

\begin{figure}[!htb]
   \centering
   \includegraphics*[width=.99\columnwidth]{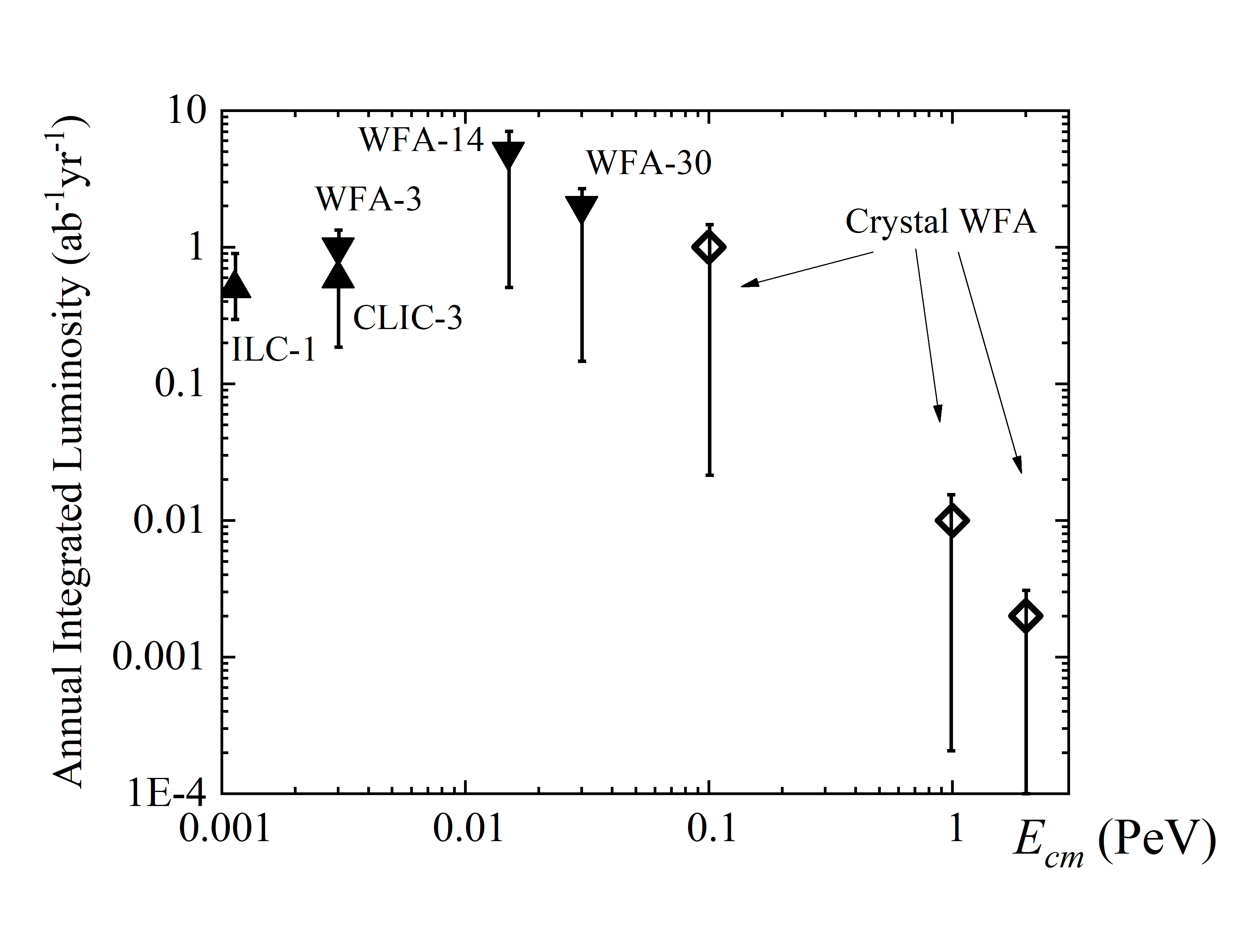}
   \caption{Estimated annual integrated luminosity for very high energy linear lepton colliders: RF-based ILC and CLIC, and plasma wakefield-based $e^+e^-$, and linear crystal wakefield $\mu^+\mu^-$.}
   \label{fig4}
\end{figure}

Fig.\ref{fig4} presents estimated luminosities of very high-energy linear lepton colliders, starting with the 1 TeV ILC and 3 TeV CLIC, and followed by wakefield acceleration (WFA) 0.01-0.03 PeV LCs based on gaseous plasma, and up to 1 PeV crystal muon LC options. 

\section{Conclusion} 


The future of particle physics is critically dependent on the feasibility of future energy-frontier colliders. The concept of feasibility is complex and includes at least three factors: feasibility of energy,luminosity, cost and construction time. This article has presented major beam-physics limits of ultimate accelerators and take a look into the ultimate energy reach of possible future colliders. A paradigm change for high-energy particle physics research is looming, as the thrust for higher energies by necessity will mean lower luminosity.

The above considerations of ultimate high-energy colliders for particle physics indicate that their major thrust is attainment of the highest possible energy $E_{cm}$, while the accelerator design challenge is high luminosity $\cal L$ and the major limit is the cost $C_c$. The cost is 
critically dependent on acceleration technology used to reach the required $E_{cm}$. The limits on $E_{cm}$ were assumed to be the total facility construction cost being less than three times the cost of the world's most powerful collider to date, the LHC, i.e., $C_c \leq 3$ LHCU. The cost limitations are not well defined, being dependent on such societal factors as the priority and availability of resources to support fundamental research. Consequently, if the affordable collider cost limit can be increased, say, 3-fold to $C_c\sim$10LHCU, that would also push the maximum collider energy $E_{cm}$ by a factor of 3-10, according to Eq.(\ref{ModelCost}). Notably, employment of already existing injectors and infrastructure can greatly help to reduce $C_c$. 

For most collider types, the pursuit of high energy typically results in low(er) luminosity. So, e.g., more than $O$(1 ab$^{-1}$/yr) at $E_{cm} \ge $ 30 TeV to 1 PeV can not be expected. 
The luminosity calculations might be assumed to be limited by the total facility (and, therefore, the beam) annual power consumption to $\sim$3 TWh/yr, again, depending on the societal priorities and considerations of ecological footprint and energy efficiency. 

For the collider types considered the following conclusions could be drawn: i) for circular $pp$ colliders the overall feasibility limit is close to or below 100 TeV ($\sim$14 TeV c.m.e. for constituents); ii) for circular $ee$ colliders the limit is at $\sim$0.5 TeV; iii) for circular $\mu \mu$ colliders the limit is about 30 TeV; iv) for linear RF-based lepton colliders and plasma $ee$/$\gamma \gamma$ colliders, the limit is between 3 and 10 TeV; v) there are exotic schemes, such as crystal channeling muon colliders, which potentially offer 100 TeV-1 PeV c.m.e., though at very low luminosity. All in all, muons seem to be the particles of choice for future ultimate HEP colliders.

\section{Acknowledgements and Further Reading}
\label{furtherreading}

This paper is mostly based on the author's presentation at the workshop on the ``Physics Limits of Ultimate Beams" \cite{PLUB} (January 22, 2021; on-line) and recent review \cite{shiltsev2021modern}. The author greatly appreciates input from and very helpful discussion on the subject of this paper with Mei Bai, William Barletta, Steve Gourlay, Vladimir Kashikhin, Valery Lebedev, Mark Palmer, Tor Raubenheimer, Thomas Roser, Daniel Schulte, John Seeman, Toshiki Tajima and Alexander Zlobin. Special thanks go to my long-term collaborator and co-author Frank Zimmermann, who always inspired me with his contributions to and visionary analysis of future colliders and suggested writing this article. 

For those who want to read more deeply on the topics touched in this article,  one can recommend the following sources:
\begin{itemize}
    \item reviews of the modern and future particle colliders -- Ref. \cite{shiltsev2021modern, myers2013accelcollidersbook, bruning2016challengesXXIbook},
    \item introductory textbooks on high-energy particle physics -- Refs. \cite{perkins200hepbook, barger2018collphysbook},
    \item on the history of accelerators and colliders -- Refs.\cite{livingston1954book, hoddeson1997SMbook, sessler2014enginesbook},
    \item on accelerator and collider technologies and costs -- Refs.\cite{seryi2016unifyingbook, zlobin2019nb3sn, shiltsev2014,florio2020economics, koizumi2020evolution, ITF}, 
    \item other publications on the topics of limits of particle colliders -- Refs.\cite{shiltsev2012ufn, zimmermann2018future, shiltsev2019ultimate}. 
\end{itemize}

{\bf {\Large Glossary}}
\\

{\bf beam-beam effects:} a variety of usually detrimental effects arising during collision of dense charged-particle bunches, such as blow-ups of beam sizes, increase of the energy spread, growth of the beam halo and particle losses; most prominent in collisions of high-intensity, high-brightness bunches. 

{\bf beam cooling:} reduction of the beam emittance (phase-space area) without loss of intensity; there are several methods for such improvement of the particle-beam quality (each with its own limits of applicability) -- radiation damping, electron cooling, stochastic cooling, laser cooling, ionization cooling, etc.

{\bf beamstrahlung:} particle's energy loss due to radiation of photons or gamma quanta, or $e^+e^-$ pair production in the strong electromagnetic fields of the opposite bunch,  one of the {\it beam-beam effects}, usually most pronounced in high-energy, high-intensity electron-positron colliders of all types.

{\bf ERL (energy-recovery linac):} power-efficient type of accelerator combining {\it linac} and storage ring; it is based on the recirculation of a charged particle beam which is first accelerated in the linac (it borrows energy from the electric fields of RF cavities), then travels through the recirculating arc before being decelerated in the same linac structure (returning the energy).

{\bf gamma(s) $\gamma$:} unfortunately, the particle physics nomenclature has the same Greek letter for gamma particles (photons), and for the relativistic Lorentz factor $\gamma=E/mC^2$; in this paper, the context is meant to make it clear which of the two meanings is being discussed.

{\bf intrabeam scattering (IBS):} is a single-beam effect caused by collisions between particles in circular accelerators; it leads to an increase in the beam emittance (size), typically occurring slowly in one or all three dimensions. This effect is most prominent in high-intensity, high-brightness bunches.

{\bf linac:} linear accelerator. 

{\bf "$m$"-poles (dipoles, quadrupoles, sextupoles, etc):} types of most commonly used accelerator magnets which have 2, 4, 6 etc poles and generate corresponding types of the magnetic fields configurations that are needed for charged-particle guidance (bending and focusing). 

{\bf RF (radio-frequency):} This is a general term for accelerator components such as cavities and structures, as well as systems like generators and controls, that provide alternating electric or electromagnetic fields at radio frequencies ranging from dozens of kHz (rarely) and MHz, to GHz and (rarely) THz 

{\bf space-charge effects:} single-beam phenomena caused by the electromagnetic interaction of particles which usually repel each other; that leads to blow-ups of beam sizes and particle losses; most prominent in low-energy (non-relativistic), high-intensity, high-brightness beams. 

{\bf Standard Model of particle physics:}  self-consistent theory describing three of the four known fundamental forces in the universe and classifying all known elementary particles; correspondingly, the eagerly sought new physics phenomena in HEP are often called {\it Beyond the Standard Model} (BSM).

{\bf synchrotron radiation:} is the electromagnetic radiation emitted when charged particles travel in curved paths; results in particle energy loss and is most pronounced in high-energy electron accelerators. 

{\bf Wake-field acceleration:} relatively novel methods of excitation of high-gradient electric fields (needed for acceleration of charged particles) by very short intense driving pulses of either lasers or electrons; such fields follow the driver pulses propagating either in plasma or in metallic/dielectric open-aperture structures - therefore, the {\it wakes}.

\newpage



\addcontentsline{toc}{section}{Bibliography}

\bibliographystyle{apalike}
\bibliography{bibliography.bib}
\end{document}